\documentclass[conference]{IEEEtran}
\IEEEoverridecommandlockouts
\usepackage{cite}
\usepackage{amsmath,amssymb,amsfonts}
\usepackage{algorithmic}
\usepackage{graphicx}
\usepackage{textcomp}
\usepackage{xcolor}
\usepackage{hyperref}
\usepackage{setspace}
\usepackage{float, longtable}
\usepackage[english]{babel}
\usepackage{mathtools}
\usepackage{amssymb}
\usepackage{adjustbox}
\usepackage{rotating}
\usepackage{makecell}
\usepackage{changepage}
\usepackage{lscape}

\def\BibTeX{{\rm B\kern-.05em{\sc i\kern-.025em b}\kern-.08em
    T\kern-.1667em\lower.7ex\hbox{E}\kern-.125emX}}

\makeatletter
\def\@IEEEpubidpullup{8\baselineskip}
\makeatother

\usepackage{fancyhdr}
\usepackage{kantlipsum}
\fancypagestyle{plain}{
\fancyhf{}
\fancyhead[C]{Conference on \LaTeX} 

}
\usepackage{eso-pic}

\begin{document}



\title{Exploring the Role of Network Centrality in Player Selection: A Case Study of Pakistan Super League\\


\thanks{*These authors contributed equally to this work.}
\thanks{$^{1}$Corresponding author.}
\thanks{A. Khan, S. J. Alam, and Q. Pasta are with the Dhanani School of Science and Engineering, Habib University, Karachi, Pakistan (e-mail: abeer.khan2048@gmail.com, sj.alam@sse.habib.edu.pk, \newline qasim.pasta@sse.habib.edu.pk).}
\thanks{M. H. Samiwala and A. Zawar are with the School of Arts, Humanities, and Social Sciences, Habib University, Karachi, Pakistan (e-mail: mariasamiwala@gmail.com, abeehazawar@gmail.com).}
}

\author{Abeer Khan*, Maria Hunaid Samiwala*, Abeeha Zawar$^{1}$, Qasim Pasta$^{1}$ and Shah Jamal Alam$^{1}$}

\maketitle

\begin{abstract}
Cricket, a popular bat-and-ball game in South Asia, is played between two 11-player teams. The Pakistan Super League (PSL) is a commercial T20 domestic league comprised of six franchise-owned teams, where player selection is competitive. In this study, an existing role-based ranking structure is assessed that evaluates player performance in the context of team belongingness to generate optimal Pakistan cricket teams for international tournaments. The underlying assumption is that since cricket is fundamentally a team sport, the performance of players compared to their peers plays a crucial role in their selection. To accomplish this, a network is generated using ball-by-ball data from previous PSL matches (2016-2022), and social network analysis (SNA) techniques such as centrality and clustering coefficient measures, are employed to quantify the level of belongingness among Pakistani cricket players within the PSL network. Characteristic network models, such as the Erdös-Rényi, Watts-Strogatz, and Barabási-Albert models are utilized to gain insights into the small-world properties of the network. By ranking players using centrality and clustering coefficient metrics, four teams are formulated, and these teams are subsequently compared to the official squad selected by the Pakistan Cricket Board (PCB) for the recent ICC Men's T20 World Cup in 2022. This evaluation sheds light on the allegations of nepotism and favoritism in team formations that have been attributed to the PCB over the years. Based on our findings, out of the 18 players in the World Cup squad, 11 were included in the teams we formed. While most of the 7 players who were not included in our teams were still selected for the ICC Men's T20 World Cup 2022, they ranked highly in our rankings, suggesting their potential and competence.
\end{abstract}


\section{Introduction}
\label{sec1}
When it comes to team formations in the realm of sports, the significance of individual player performance is of utmost importance. However, effective team dynamics and player-to-player interactions are also pivotal to success in numerous popular sports, such as basketball, football, hockey, etc. Cricket, a bat-and-ball sport played on a pitch by two teams of eleven players is no exception. In cricket, the batting team aims to score as many runs as possible in a given number of overs, while the opposing bowling team strives to secure wickets to restrict the batting team's runs. It is the most commercialized sport in Pakistan and the Pakistan Super League (PSL) stands as the most renowned domestic T20 cricket league in the country, attracting substantial revenue amounting to billions of Pakistani rupees in its recent seasons \cite{b9, b10}. The League is held annually during the Spring season and currently features six franchise teams that represent various Pakistani cities. Players are drafted by different franchises before the commencement of a new season because of which some players may find themselves switching teams multiple times throughout their PSL career, while others may remain with a single franchise throughout \cite{b12}. 

Pakistan's cricket industry has faced allegations regarding the prevalence of nepotism and favoritism in player selection for team formations \cite{b6}, with prominent cricket figures (such as Wasim Akram and Shoaib Malik) voicing their concerns about how personal connections often take precedence over talent and hard work \cite{b7, b8}. These allegations have raised concerns about the transparency and fairness of the player selection process, calling into question the principles of meritocracy and equal opportunity.

In this study, we assess the implementation of a role-based ranking structure based on player performance evaluation, as proposed by Dey et al. \cite{b1}. This approach is rooted in the belief that cricket is fundamentally a team sport, where the performance of players compared to their peers plays a crucial role in their selection. Social Network Analysis (SNA) helps in understanding social structures and interactions, and prior approaches have successfully utilized it to gain crucial insights into the network dynamics of various sports teams. Therefore, we aim to quantitatively measure the degree of belongingness of individual players within a team, particularly within the context of Pakistani cricket, an area that has received limited attention in previous research. By analyzing the network dynamics of the PSL and employing network centrality measures as a quantitative lens, we evaluated the significance of individual players based on their performance within the team network. Using this approach, we formulated and assessed teams alongside the official squad chosen by the Pakistan Cricket Board (PCB, the country’s official cricketing body) for the recent ICC Men’s T20 World Cup 2022. This work is a step towards testing allegations surrounding nepotism and favoritism in team formations by the PCB. Moreover, the findings of this research could provide valuable insights for PCB's stakeholders and administrators, aiding them in making informed decisions regarding optimal team formations for future tournaments.

The rest of the paper is organized as follows. Section 2 discusses the \hyperref[sec2]{Related Work}, providing insights into relevant existing research. In Section 3, the \hyperref[sec3]{Methodology} explains the data acquisition process and research procedures employed. Section 4 presents the \hyperref[sec4]{Network Analysis} utilizing various centrality measures and identifying network characteristics through comparison with characteristic network models. Section 5 delves into the \hyperref[sec5]{Team Formations} obtained, and evaluates the effectiveness of the approach by a comparison with the selected squad of the last ICC Men's T20 World Cup in 2022. Lastly, Section 6 presents the \hyperref[sec6]{Conclusion \& Future Improvements}, offering insights into the outlook and next possible steps.

\section{Related Work}
\label{sec2}
Several studies have demonstrated the potential of social network analysis in understanding various types of interactions within collective systems and their implications. For instance, Mukherjee \cite{b2} conducted a study utilizing network approaches to investigate the interaction among batters in international cricket matches. By constructing batting partnership networks (BPNs) and analyzing network metrics, he observed small-world behavior within the networks when comparing these metrics to the Erdös-Rényi framework. Notably, the findings also indicated that the most connected batter did not necessarily hold the most important or influential position, and players with the highest batting averages did not always occupy the most central positions. Furthermore, he investigated the structure of BPNs to determine players' roles and expanded his methodology to quantify the efficiency, performance, relevance, and impact of removing a member from the team using different centrality values, thereby providing a more comprehensive understanding of individual contributions within a team. Mukherjee \cite{b3} then extended the application of network analysis in cricket by evaluating players' performance to create teams for international matches. While batting and bowling averages conventionally rank players, he acknowledged the importance of considering the technique and context in which runs are scored, or wickets are taken. By using SNA on player-vs-player data, he constructed a directed and weighted network of batters and bowlers, emphasizing the effectiveness of SNA in evaluating individual performances within a team.

Dey et al. \cite{b1} employed small-world network features to assess and construct cricket teams that can demonstrate optimal performance. Using T20 cricket (2014-2016) matches as their data set, they represented players as network nodes and interactions between team members as edges to build a bidirectional weighted network. Using measures of network centrality, a ranking structure was constructed that would then be used to identify the top-ranking players for team selections. This approach provided a novel method for evaluating team belongingness and individual performance and generated ratings to construct a team based on player performance, which was then validated against the 2016 IPL teams.

Concerning other sports domains, Passos et al. \cite{b4} aimed to explore the applicability of network analysis in capturing and studying collective behaviors within a water polo team. They revealed a wide range of synchronization and coordination patterns that differentiated successful from unsuccessful performance outcomes. Consequently, their findings provided compelling evidence for the effectiveness of network analysis as a quantitative approach to representing the interactions that arise within a team. Similarly, Wiig et al. \cite{b5} explored network analysis in football to identify important players, and analyzed patterns of movement and distribution within teams. By constructing networks based on player passes and different centrality metrics, they investigated alternative approaches for rating player passes. Their findings, based on data from Norway's top division over four seasons, demonstrated that considering multiple measures and network weights allowed for identifying crucial passes across different phases of play and positions on the field.

\section{Methodology}
\label{sec3}
\begin{figure}[!t]
    \centering
    \includegraphics[width=3.2in]{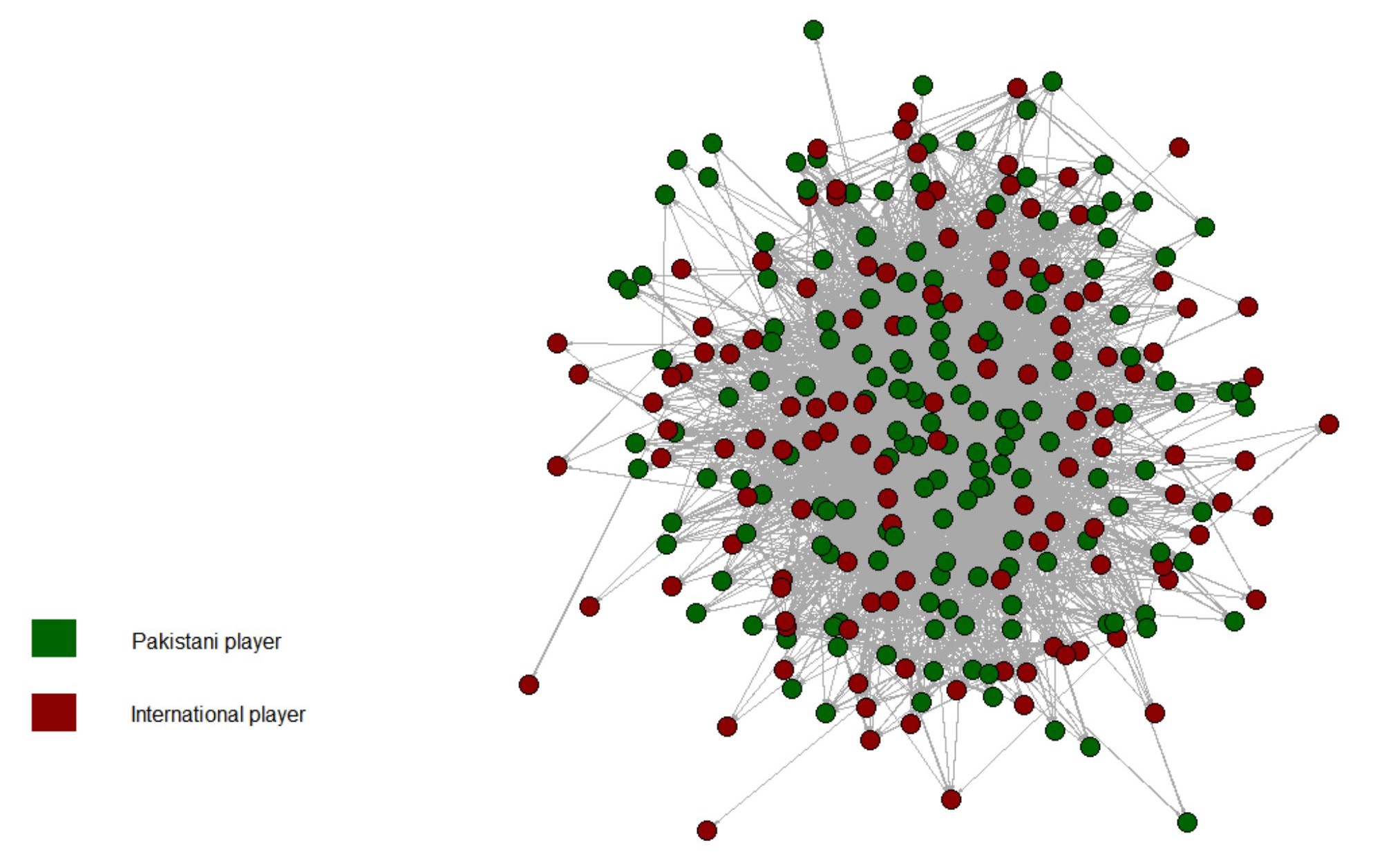}
    \caption{The constructed PSL T20 Player-to-Player interaction network showing domestic and international players}
    \label{fig1}
\end{figure}

The first season of the Pakistan Super League (PSL) took place in 2016, and until 2022, a total of seven consecutive seasons have been played. To obtain detailed ball-by-ball data for each season, Cricsheet was utilized as a source, providing comprehensive data for a wide range of cricket matches \cite{b13}. Specifically, the JSON data sets obtained contained ball-by-ball data for 214 PSL cricket matches that occurred between 2016 and 2022. Python scripts were developed to extract valuable insights regarding player participation in each match that had occurred. We identified all the players who participated in the league seasonally, along with the teams they played for and the total number of matches played. The extracted data served as the foundation for the bi-directional weighted network in Figure \ref{fig1}, where each team member was represented as a node, and player-to-player interactions were represented as edges. The weighted edges are created from the ratio of common matches and the total number of matches played by each player. It is important to note that our network formation encompassed all PSL players, including retired and international players. The libraries utilized in R for the generation of the network were igraph, sqldf, DirectedClustering, reshape2, and dplyr.

Due to the extensive volume of data, manual verification of the data set was not feasible. However, we discovered that the data set lacked information for a match between Peshawar Zalmi and Karachi Kings, which occurred on February 11, 2016. To address this, we manually gathered ball-by-ball information for the missing match using ESPN Cricinfo \cite{b15}. Additionally, a meticulous examination of player information was necessary as we encountered numerous discrepancies within the acquired data set. For the purpose of verification of player identification, Cricsheet maintains a registry comprising CSV files that include entries of individuals associated with the cricket industry \cite{b14}. These files incorporate unique identifiers from sources like ESPN Cricinfo \cite{b15}. To ensure the precision of our analysis, we identified and corrected instances of duplicate entries and multiple players sharing the same names, utilizing the unique identifiers provided by the registry.

The generated network is subjected to a comparative analysis with several prominent network models, namely the Erdős-Rényi (E-R), the Watts-Strogatz (W-S), and the Barabási-Albert (B-A) models \cite{b16, b17, b18}. The objective of this comparison is to determine and analyze the network's characteristics. To quantify and evaluate a player's sense of belongingness within the network, we employ centrality measures such as degree centrality, betweenness centrality, and closeness centrality, along with a local clustering coefficient. Using these measures, a role-based ranking structure is constructed, exclusively focusing on Pakistani players participating in the Pakistan Super League (PSL) who were eligible for the ICC T20 Men’s World Cup 2022. We implemented a series of eligibility constraints for selecting players for the team selection. We attempt to assemble a playing XI team that exhibits the highest potential for success by selecting players who demonstrate exceptional team-playing abilities. 

\section{Network Analysis}
\label{sec4}

\subsection{Small World Characteristics \& Comparative Analysis}

\begin{figure}[!t]       
    \includegraphics[width=1.45in]{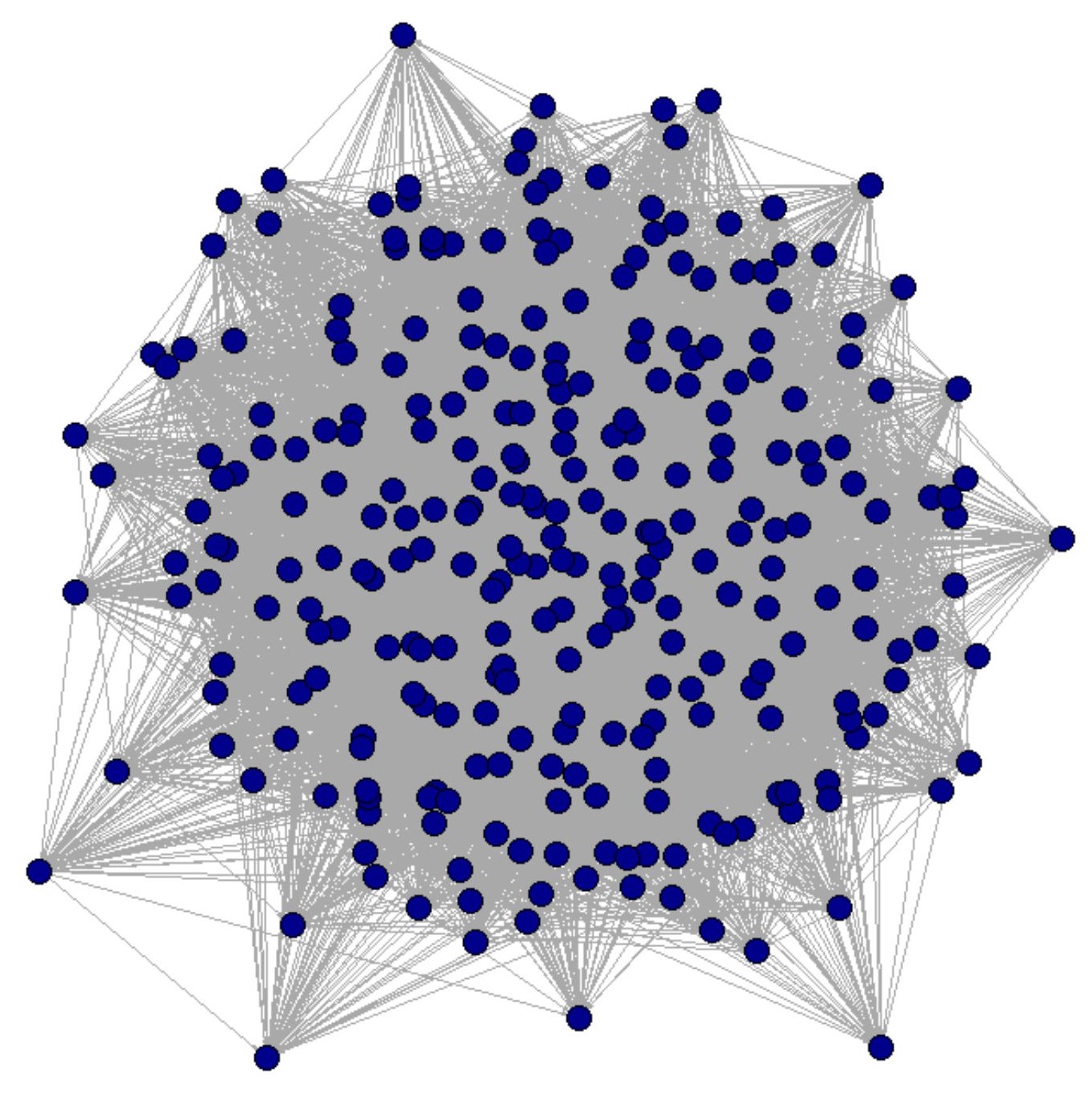}  
    \includegraphics[width=1.45in]{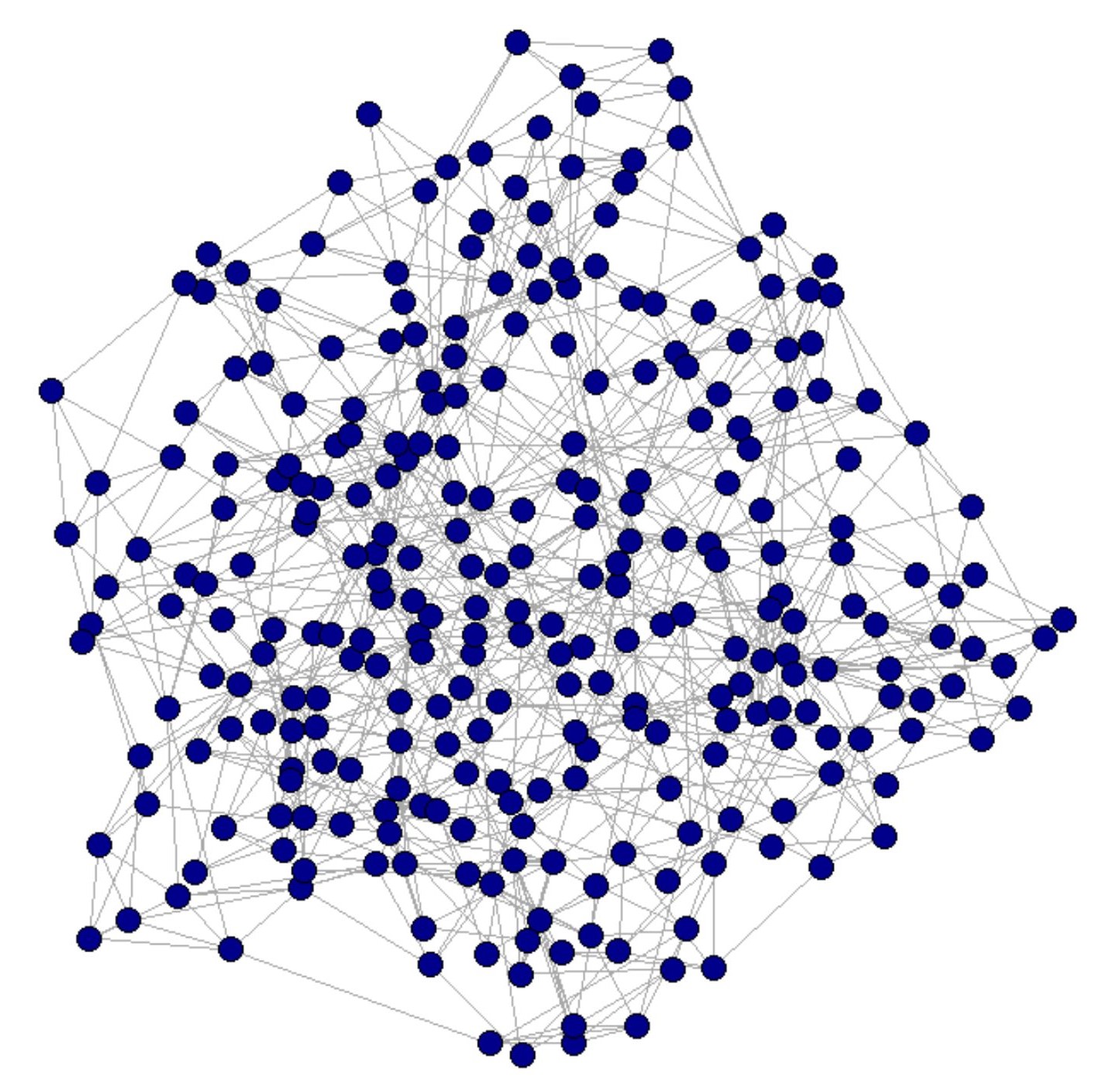}
    \centering \includegraphics[width=1.45in]{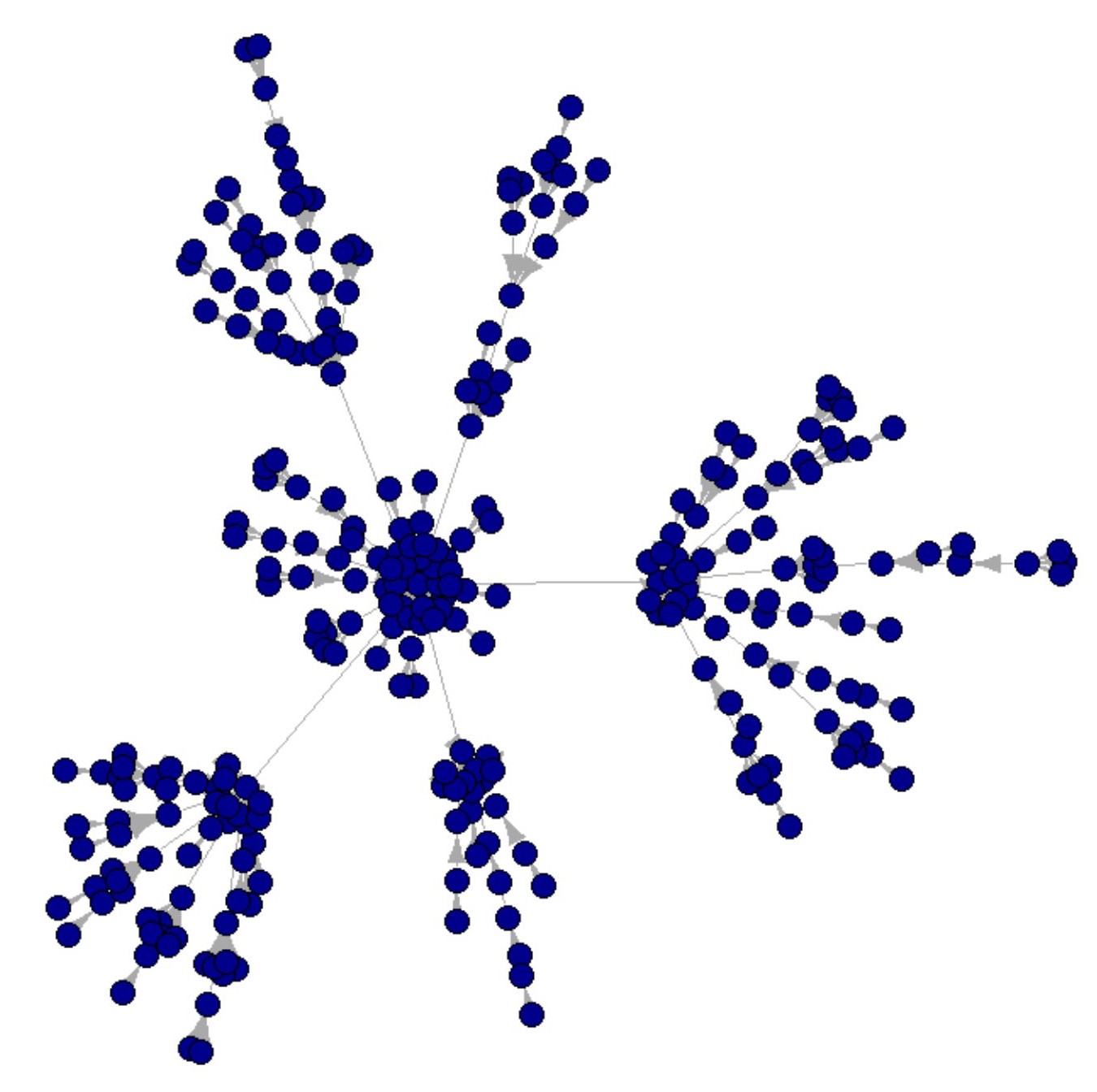}
    \caption{(a) PSL Erdős-Renyi Network (b) PSL Watts-Strogatz Network (c) PSL Barabási-Albert Network}
    \label{fig3}
\end{figure}

In a network with a small world structure, the distance between any two nodes is significantly less. For our centrality analysis, the connections between nodes hold great importance, making it crucial to assess the small-world phenomenon within the PSL network. In order to thoroughly examine the characteristics of the PSL network, encompassing seven seasons, we utilized various measures such as global clustering, average path length, and degree distribution. Additionally, we conducted a comparison with three characteristic network models; the Erdős-Renyi (E-R), Watts-Strogatz (W-S), and the Barabási-Albert (B-A) models. These networks were generated using RStudio, comprising the same number of nodes and edges as the PSL network.

The average path length of a network shows how well the nodes are connected to other nodes in the network. In the case of a small world network, this value should be significantly low, close to the natural logarithm of its number of nodes, denoted as ln(N), where N is the number of nodes. For the PSL network, which consists of 284 nodes (players), the expected average path length should be roughly 5.65 for it to qualify as a small-world network. The calculated average path length of the network is substantially lower, at 2.16. This result suggests that the PSL network exhibits characteristics of a highly interconnected small world. Essentially, it implies that the average number of hops required to connect any two players in the network is less than three, highlighting its exceptional level of connectivity. When comparing these findings to the network models, the PSL network's average path length surpasses that of the PSL E-R network (1.96) but falls short of the PSL W-S network (4.02). Notably, it closely aligns with the average path length of the PSL B-A network (2.08), indicating a potential preference for players to selectively form connections with other players.

\begin{figure}[!t]       
    \includegraphics[width=3.30in]{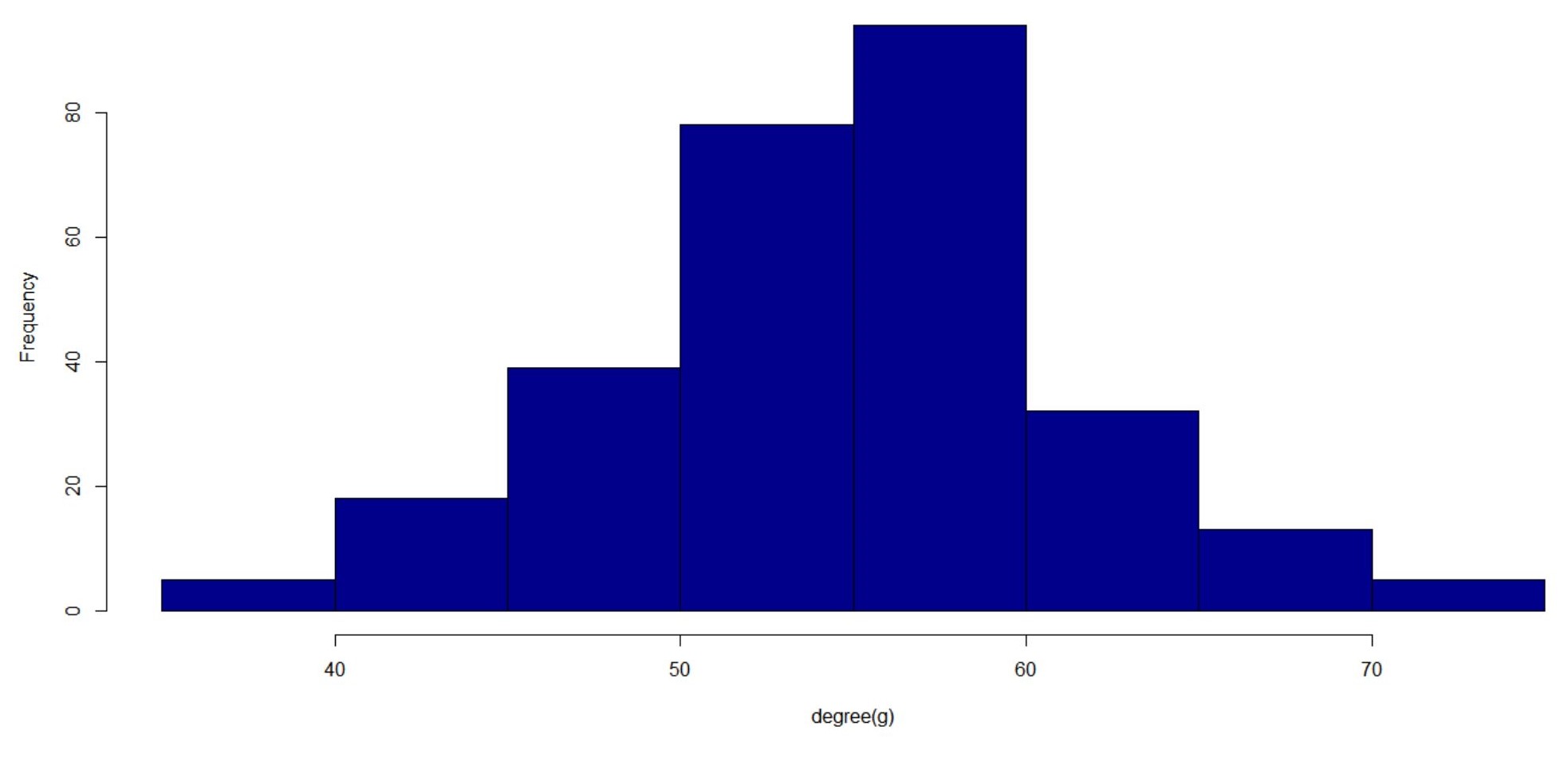}
    \caption{Degree Distribution of Erdos-Renyi PSL Network}
    \label{fig4}
\end{figure}

\begin{figure}[!t]       
    \includegraphics[width=3.30in]{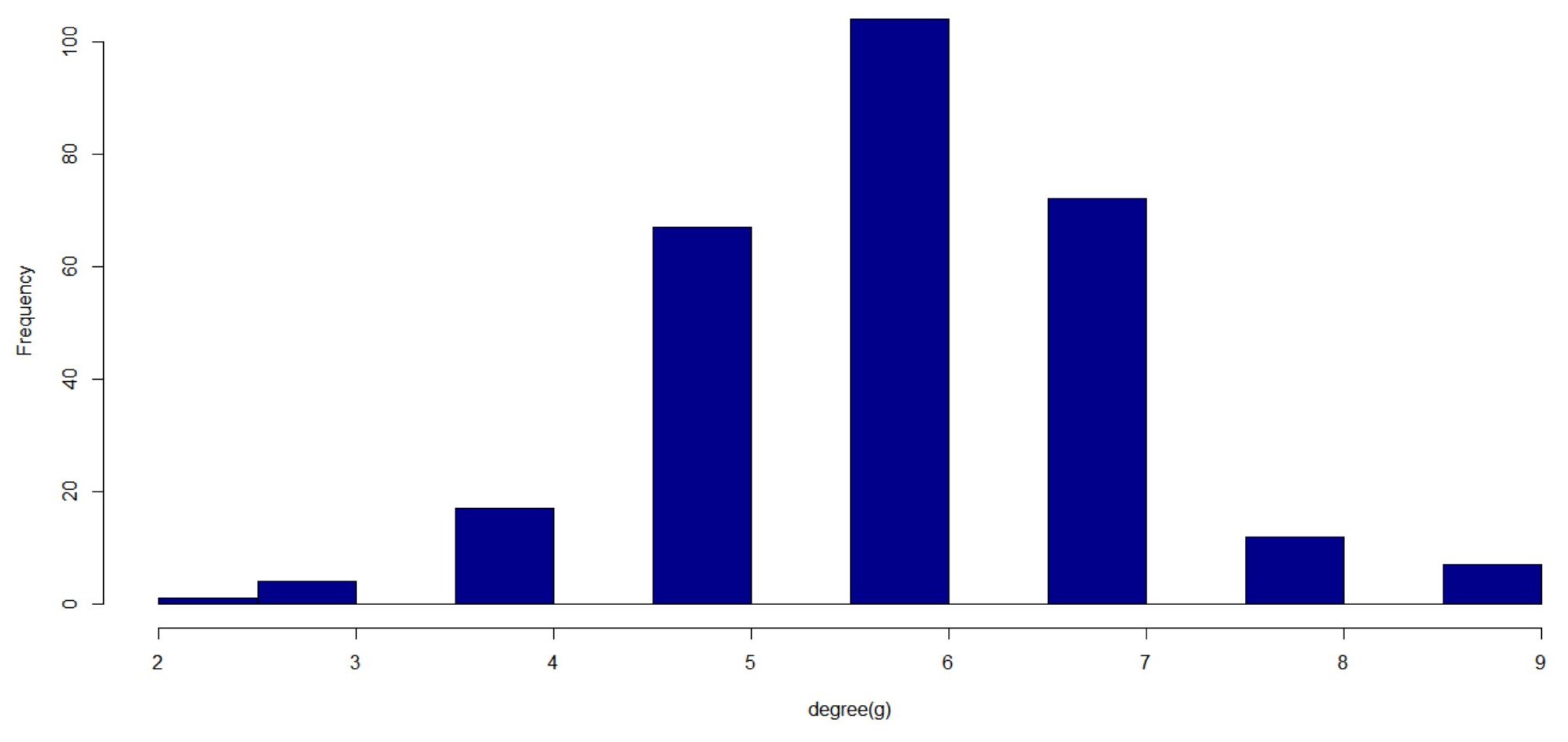}
    \caption{Degree Distribution of Watts-Strogatz PSL Network}
    \label{fig5}
\end{figure}

\begin{figure}[!t]       
    \includegraphics[width=3.30in]{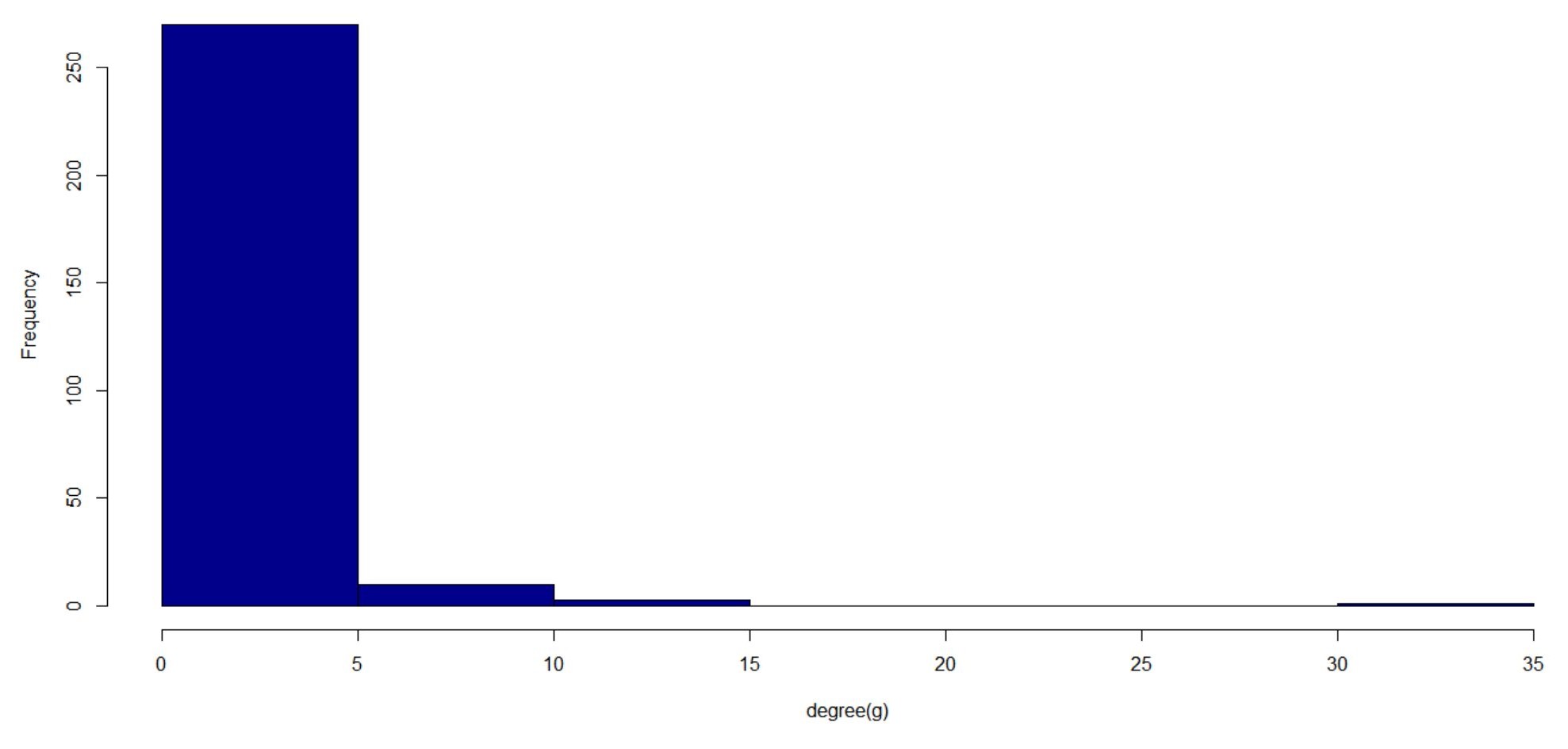}
    \caption{Degree Distribution of Barabasi-Albert PSL Network}
    \label{fig6}
\end{figure}

\begin{figure}[!t]       
    \includegraphics[width=3.30in]{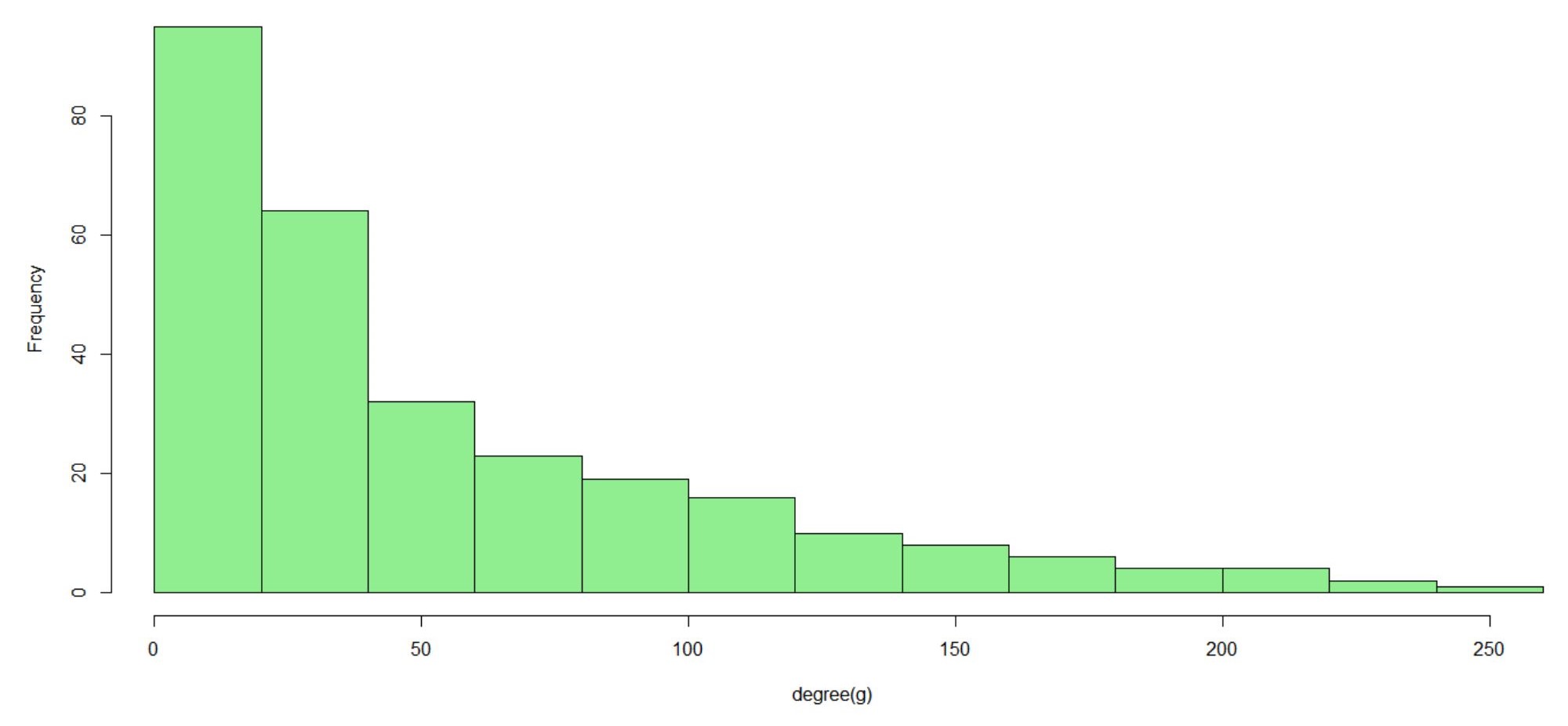}  
    \caption{Degree distribution of PSL network}
    \label{fig7}
\end{figure}

The phenomenon of preferential attachment is further supported by examining the degree distribution across all networks. In the case of the PSL network, its degree distribution exhibits a skewed pattern, aligning with both the Power Law distribution and the degree distribution of the PSL B-A network, as depicted in Figures \ref{fig6}, and \ref{fig7}. This starkly contrasts with the degree distributions observed in the PSL E-R and PSL W-S networks, depicted in Figures \ref{fig4}, and \ref{fig5}. This discrepancy indicates the presence of hubs in the PSL network, according to Barabási-Albert's network model theory, which suggests that new nodes in a network consciously attach themselves to existing nodes with higher degrees. In the context of the PSL network, this implies the existence of a small number of players who are highly interconnected, while the majority have relatively fewer connections.

Furthermore, the global clustering of the PSL network was found to be 25.40\%, which is closest to the average clustering coefficient of 33.00\% observed in the PSL W-S network. It significantly exceeds the clustering coefficient of 9.74\% for the PSL E-R network and the 0\% clustering coefficient for the PSL B-A network. This finding contradicts the preferential attachment pattern observed in the path length and degree distribution analyses of the PSL network. However, this discrepancy can be explained by considering that the network models are generated randomly, whereas the players in the actual PSL network engage in intentional interactions based on the matches they play together. Consequently, the presence of clusters within the PSL network indicates groups of players who have participated in multiple matches together, while hubs signify players who consistently secured positions in their team's starting lineup, were rarely injured, or frequently transferred between teams.

\subsection{Degree Centrality}
Degree centrality is a measure that quantifies the number of direct connections, or 'one hop' relationships, that a node has with other nodes in a network. We employed degree centrality to analyze various aspects, such as popular players, and players who potentially possess a significant amount of information or have the ability to quickly connect with the broader network. Additionally, we examined both in-degree (number of inbound links) and out-degree (number of outbound links). We identified a maximum degree centrality value of 252. This signifies that the player Wahab Riaz has directly interacted, or played matches, with 252 other players throughout his PSL career. On average, the degree centrality value across all players is 55.21, indicating that, on average, a player 'n' has engaged in direct matches with 55 other players during their PSL career.

The degree distribution of the PSL network demonstrates a skewed pattern that aligns with the B-A model or Power Law distribution. Specifically, 95 nodes within the network possess a degree centrality of 20 or less (ranging from 0 to 20), while only 1 node has a degree exceeding 240. This observation suggests the existence of hubs in the PSL network, indicating that there are only a few players who exhibit exceptionally high levels of connectivity, while the majority of players have comparatively lower degrees of connection. The presence of hubs can indicate players that were frequently in the starting eleven of their team, which can be an acknowledgment of their ability and their experience. 
\subsection{Betweenness Centrality}
\begin{figure}[!t]       
    \includegraphics[width=3.30in]{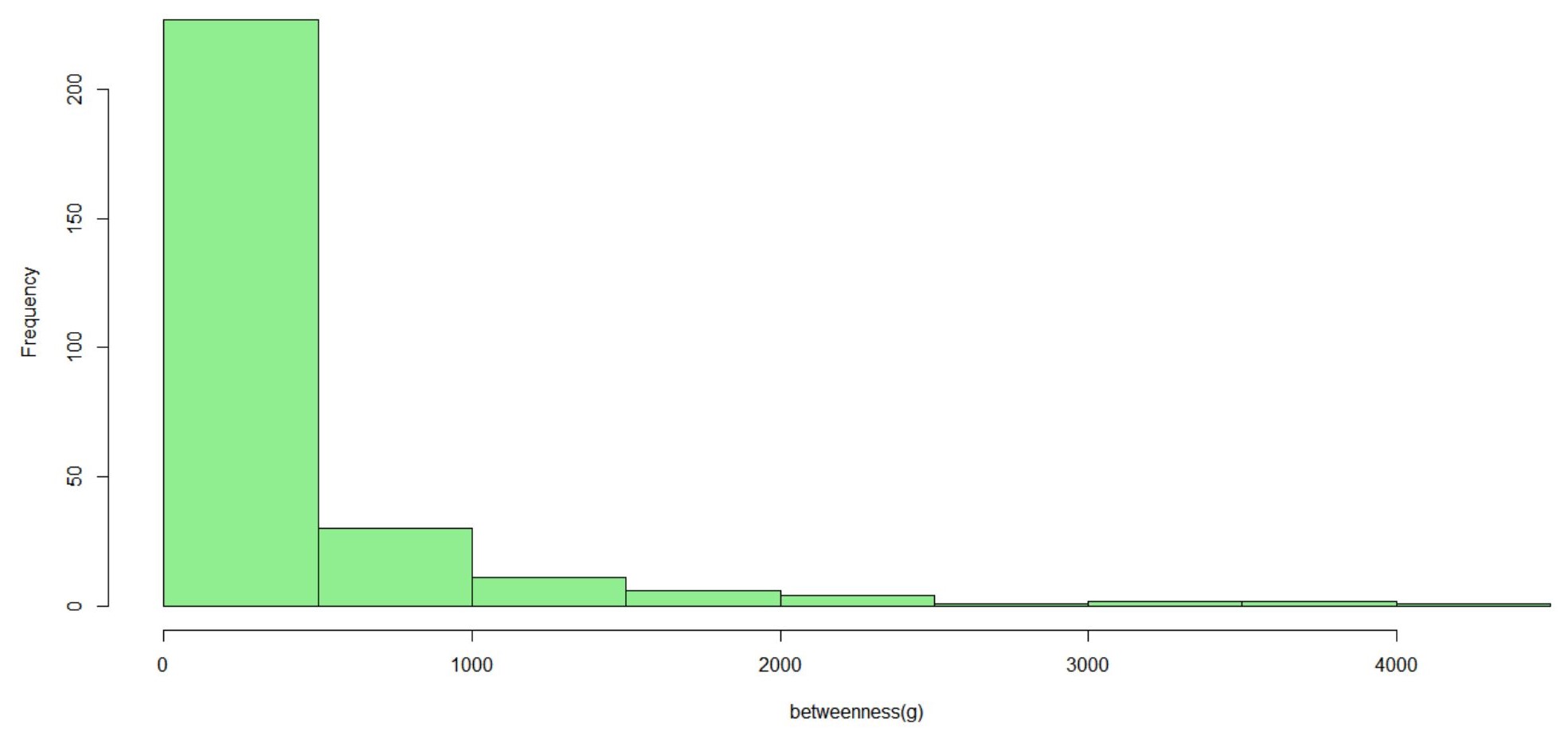}
    \caption{Betweenness centrality distribution of PSL network}
    \label{fig8}
\end{figure}
Betweenness centrality for a node 'n' in a network quantifies the proportion of total shortest paths that pass through that node 'n' relative to all possible shortest paths within the network. A player with a high betweenness centrality holds significant influence over other players, as they are dependent on passing through the player to interact with other players the player is connected to. The average betweenness centrality of the PSL network is 328. This suggests that, on average, for a player 'n' in the PSL network, there are 328 shortest paths that include 'n' in-between, when it comes to other players trying to interact with each other. Similarly, the maximum betweenness centrality value recorded is 4423.40, while the minimum value is 0. Furthermore, the distribution of betweenness centrality in the network exhibits a pronounced skew, adhering to the Power Law, as depicted in Figure \ref{fig8}. This indicates that only a select few players possess high betweenness centrality, while the majority do not. Only 10 players have a betweenness centrality value exceeding 2000, whereas 227 players have a betweenness centrality ranging from 0 to 500. Hence, the mean value might not provide an accurate assessment of betweenness for this network. 

\subsection{Closeness Centrality}
\begin{figure}[!t]       
    \includegraphics[width=3.30in]{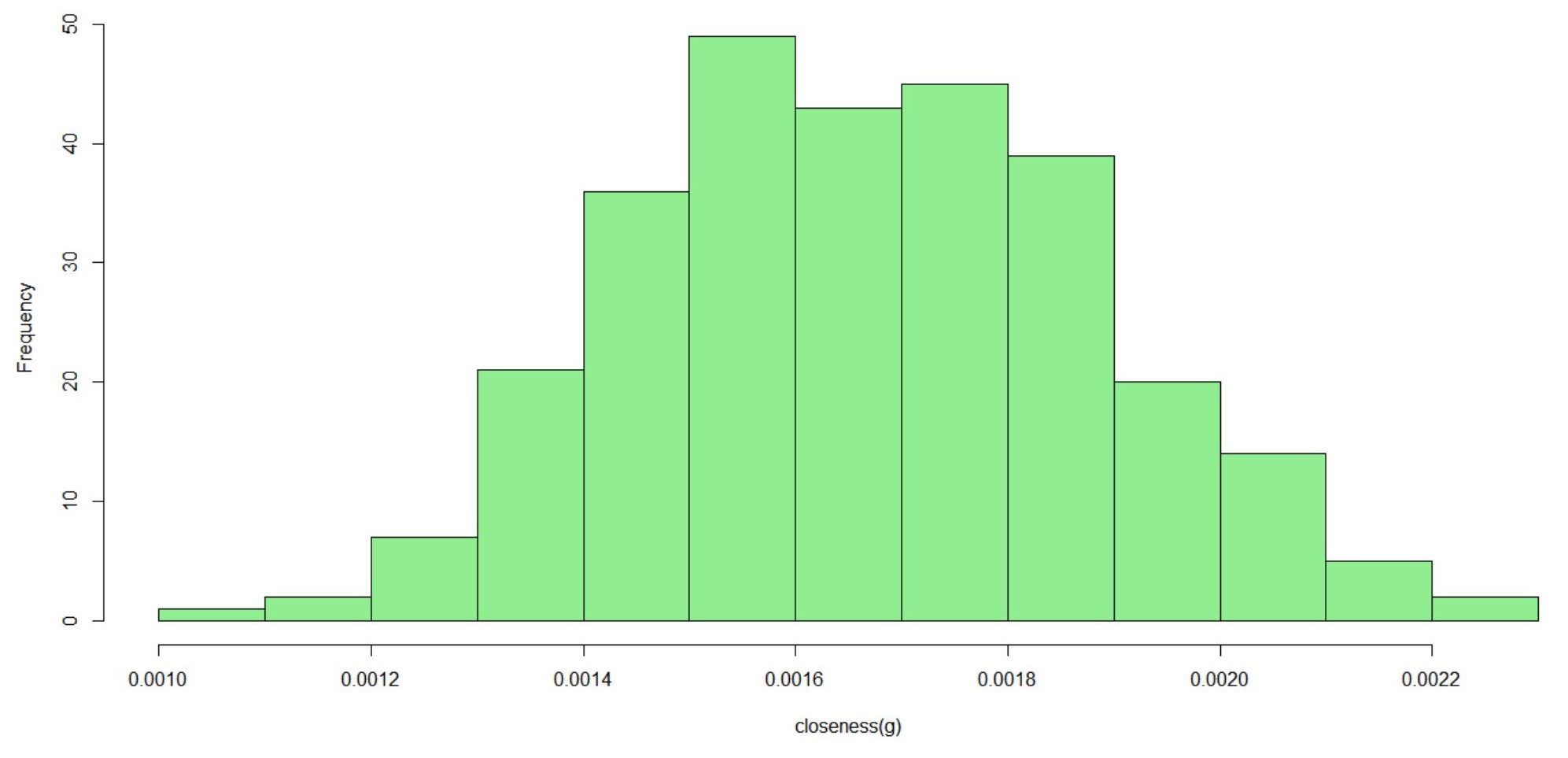}
    \caption{Closeness centrality distribution of PSL network}
    \label{fig9}
\end{figure}
Closeness centrality serves as a measure of how closely connected a node is to others in the network. It is the inverse of the distance between any two nodes in the network. A high closeness centrality value indicates that a node is well connected to other nodes in the graph. This implies that the node can quickly each other nodes.
While the clustering coefficient assesses the local connectivity of a node, the closeness centrality provides insights into its global reach. The average closeness centrality of the network is 0.0017. The distribution of closeness centrality follows a normal distribution pattern, as depicted in Figure \ref{fig9}. At the tail end, 3 nodes have a closeness centrality between 0.0010-0.0012, while 7 nodes have a closeness centrality ranging from 0.0021-0.0023. This indicates that only a limited number of nodes possess the ability to reach a very large number of other nodes. Additionally, 49 nodes have a closeness centrality ranging from 0.0015-0.0016. 

\subsection{Clustering Coefficient}
\begin{figure}[!t]       
    \includegraphics[width=3.30in]{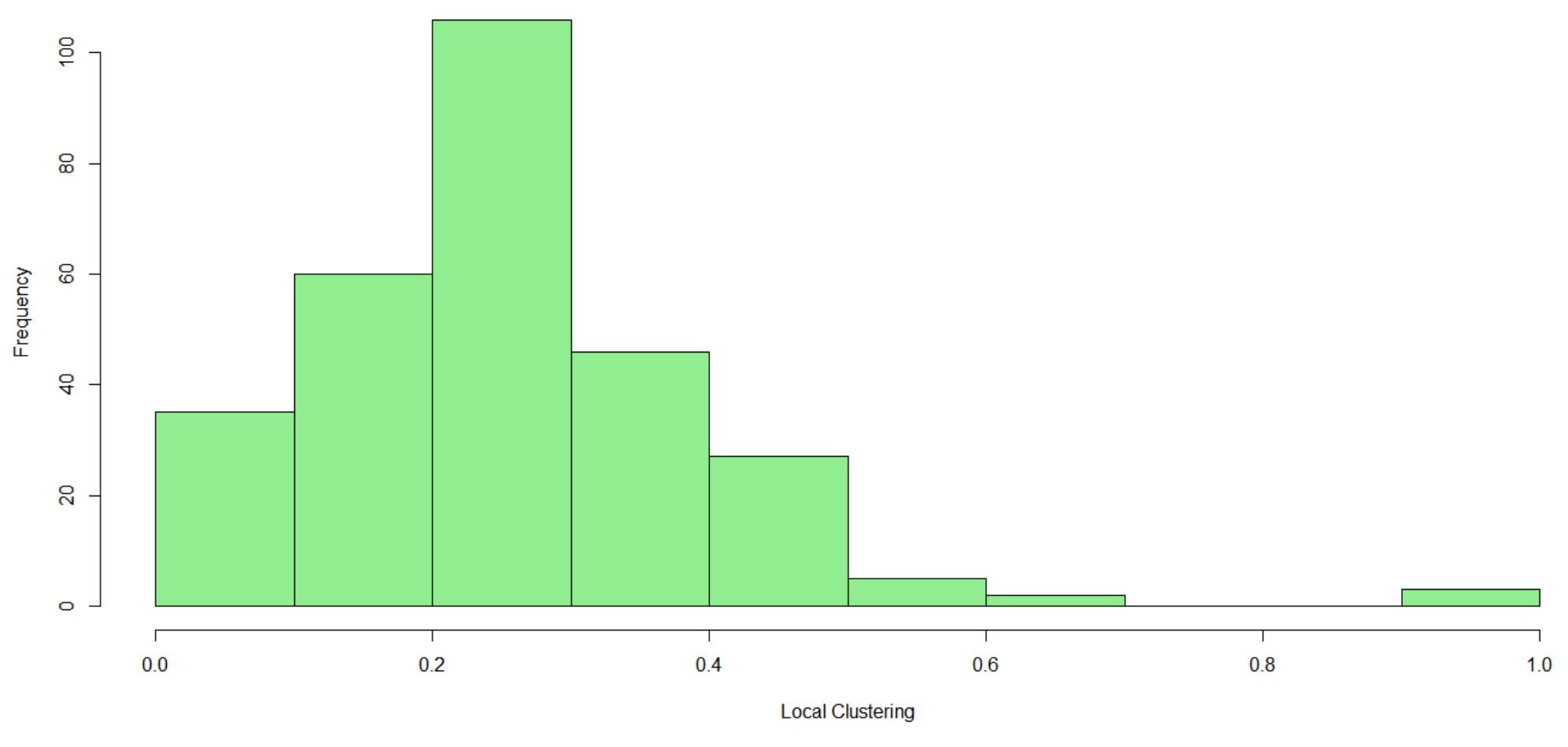}
    \caption{Clustering coefficient distribution of PSL network}
    \label{fig10}
\end{figure}

The clustering coefficient of a node within a network reflects its ability to form local clusters, representing the number of nodes influenced by a specific node. A high clustering coefficient indicates that a node's neighbors are well connected to each other. The global clustering coefficient of the PSL network is calculated to be 25.40\%, demonstrating characteristics of small-world networks as previously mentioned. This implies that, on average, 25\% of a player's ('n') neighbors are connected to each other and that the neighbors rely heavily on player 'n' to interact with each other, thereby establishing 'n' as an important player within the network. Additionally, it indicates the formation of a cluster around 'n', wherein 'n' plays a crucial role in facilitating interactions among its neighboring players.

The local clustering coefficient follows a skewed normal distribution pattern, as depicted in Figure \ref{fig10}. Only 10 players possess clustering coefficients exceeding 0.50 or 50\%, indicating that these players and their neighbors are equally dependent on each other. At the extremes of the distribution, only 3 nodes have a local clustering coefficient ranging from 0.90 to 1.00, while 35 nodes exhibit coefficients between 0.00 and 0.10. This observation suggests that while the overall distribution follows a normal pattern, there is a notable disparity in the tail ends, creating a mixture between a normal and a Power Law distribution. 

\section{Team Formations}
\label{sec5}
\subsection{Performance Pool}

\begin{figure}[!t]
    \centering
    \includegraphics[width=3.2in]{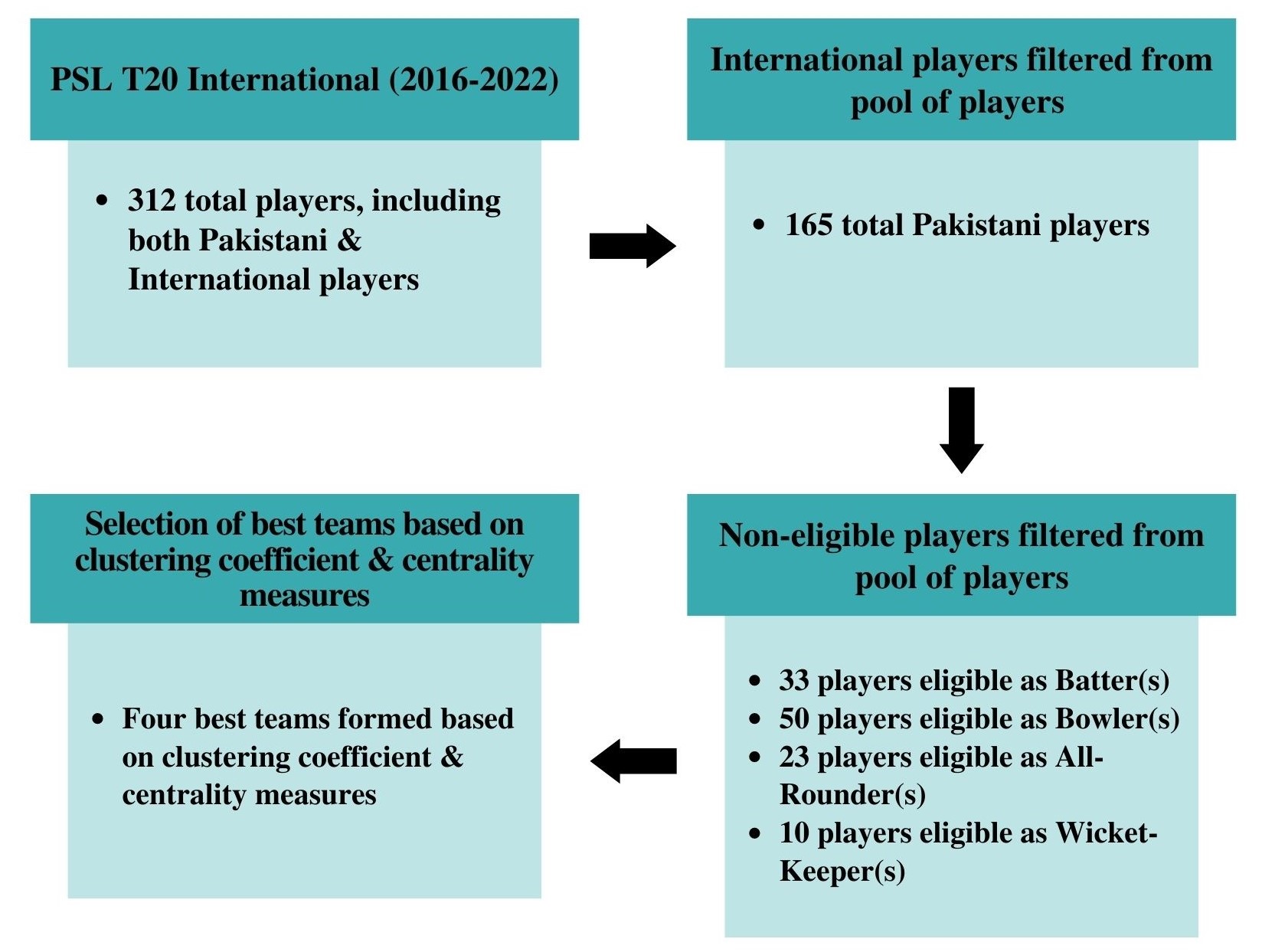}
    \caption{A flow diagram of the methodology for player selection}
    \label{fig2}
\end{figure}

We proposed a team performance strategy by implementing a performance pool selection process for Pakistani players, categorizing them as eligible or non-eligible. The eligibility and non-eligibility criteria were self-determined because some players do not formally announce their retirement, and there is no established maximum age limit for participation in ICC T20 Men's World Cup. Initially, international players were excluded from the pool. A player is deemed eligible if they have not been prohibited by the Pakistan Cricket Board (PCB) or International Cricket Council (ICC) from participating in T20 cricket, are below 40 years of age, and have not retired from international T20 cricket. Conversely, a player is considered non-eligible if they have been recently or currently prohibited by the PCB or ICC from participating in T20 cricket, are above 40 years of age, fall between the age range of 35 and 40 but have never played T20 cricket at the international level (or their last international T20 match was played five years ago), or have officially retired from international T20 cricket. We manually cross-referenced this information using ESPN Cricinfo \cite{b14} as a source of verification.  

Following Dey et al.'s methodology \cite{b1}, a minimum performance benchmark was also required in building the performance pool of players. However, in the context of PSL, the later seasons saw a substantial influx of young and new players. This presented a challenge in devising a performance benchmark, as these emerging players lacked sufficient experience compared to their more seasoned teammates. As a result, the more experienced players generally had higher overall performance, making it difficult to establish an equitable benchmark. 
    
Certain disparities were identified regarding the roles of three players in relation to their participation in the official ICC T20 team and their roles in their PSL career: i) Iftikhar Ahmed predominantly fulfilled the role of a batter in the PSL, but in the T20 Men's World Cup 2022, he was categorized as an all-rounder; ii) Mohammad Haris primarily served as a batter in the PSL, whereas in the T20 Men's World Cup 2022, he was designated as a reserve wicket-keeper; and iii) Mohammad Wasim, primarily was recognized as an all-rounder in the PSL. However, in the T20 Men's World Cup 2022, he assumed the role of a bowler. To ensure consistency in our analysis, we considered their original roles as observed in their PSL careers.

\subsection{Teams}
\begin{table*}
\begin{center}
\caption{Our optimal teams of top-ranking players with respect to centrality and clustering coefficient metrics }
\label{tab1}
\small
\begin{tabular}{|c|c|c|c|}
\hline
\rule[-1ex]{0pt}{3.5ex}Degree Distribution & Betweenness Centrality & Closeness Centrality & Clustering Coefficient \\
\hline
\rule[-1ex]{0pt}{3.2ex}Fakhar Zaman (BAT)        & Fakhar Zaman (BAT)        & Fakhar Zaman (BAT)        & Saif Badar (BAT) \\
Babar Azam (BAT)          & Babar Azam (BAT)          & Babar Azam (BAT)          & Gulraiz Sadaf (BAT)       \\
Iftikhar Ahmed (BAT)      & Iftikhar Ahmed (BAT)      & Iftikhar Ahmed (BAT)      & Umar Siddiq (BAT)         \\
Sohaib Maqsood (BAT)      & Sohaib Maqsood (BAT)      & Asif Ali (BAT)            & Sahibzada Farhan (BAT)    \\
Khushdil Shah (BAT)       & Sharjeel Khan (BAT)       & Sohaib Maqsood (BAT)      & Imam-ul-Haq (BAT)         \\
Sharjeel Khan (BAT)       & Khushdil Shah (BAT)       & Khushdil Shah (BAT)       & Muhammad Faizan (BAT)     \\
Mohammad Rizwan (WK)      & Mohammad Rizwan (WK)      & Mohammad Rizwan (WK)      & Bismillah Khan (WK)       \\
Sarfaraz Ahmed (WK)       & Sarfaraz Ahmed (WK)       & Sarfaraz Ahmed (WK)       & Gauhar Ali (WK)           \\
Imad Wasim (ALL)          & Imad Wasim (ALL)          & Imad Wasim (ALL)          & Mohammad Taha (ALL)       \\
Faheem Ashraf (ALL)       & Faheem Ashraf (ALL)       & Faheem Ashraf (ALL)       & Agha Salman (ALL)         \\
Shadab Khan (ALL)         & Shadab Khan (ALL)         & Mohammad Nawaz (ALL)      & Ahmed Safi Abdullah (ALL) \\
Wahab Riaz (BALL)         & Wahab Riaz (BALL)         & Hasan Ali (BALL)          & Mohammad Irfan (5) (BALL) \\
Hasan Ali (BALL)          & Hasan Ali (BALL)          & Wahab Riaz (BALL)         & Zafar Gohar (BALL)        \\
Shaheen Shah Afridi (ALL) & Shaheen Shah Afridi (ALL) & Shaheen Shah Afridi (ALL) & Mohammad Umar (BALL)      \\
Mohammad Irfan (BALL)     & Mohammad Irfan (BALL)     & Rumman Raees (BALL)       & Yasir Shah (BALL)         \\
Haris Rauf (BALL)         & Haris Rauf (BALL)         & Mohammad Irfan (BALL)     & Ahmed Daniyal (BALL)      \\
Rumman Raees (BALL)       & Rumman Raees (BALL)       & Haris Rauf (BALL)         & Mir Hamza (BALL)          \\
\rule[-1ex]{0pt}{1.5ex} Hassan Khan (BALL)        & Hassan Khan (BALL)        & Mohammad Hasnain (BALL)   & Arshad Iqbal (BALL) \\ 
\hline 
\multicolumn{4}{l}{\rule[-1ex]{0pt}{3.5ex}BAT: Batter; BALL: Bowler; WK: Wicket-keeper; All: All-rounder.} \\
\end{tabular}
\end{center}
\end{table*}

We created teams based on our four centrality metrics, each comprising of 18 players, ranked in descending order based on centrality, and clustering coefficient metrics. Following the structure of the official PCB squad for the ICC T20 Men's World Cup 2022, which consists of 6 batters, 3 all-rounders, 2 wicket-keepers, and 7 bowlers, we adopted the same composition when selecting players for our team formations. As depicted in Table \ref{tab1}, the team formed based on the local clustering coefficient measure differs completely from the other three teams constructed using centrality measures. Interestingly, none of the 18 players selected based on the local clustering coefficient were chosen for the ICC T20 Men's World Cup 2022. This raises concerns regarding the accuracy of the clustering coefficient as an indicator of a player's significance within the team. A high clustering coefficient suggests that all the player's neighbors are connected to each other, potentially indicating that they do not rely on that specific player to facilitate their interactions. However, comprehending the underlying reasons for this discrepancy exceeds the scope of this study. Therefore, the focus of the analysis will be on the teams formed using centrality measures, comparing them to the players selected for the ICC T20 Men's World Cup 2022 by the PCB.

Out of the 21 distinct players found in these three centrality-based teams, 11 players were part of the official Pakistan squad, while the remaining 10 players were not included. The players who did not make it to the official squad are Sohaib Maqsood, Sharjeel Khan, Sarfaraz Ahmed, Imad Wasim, Faheem Ashraf, Wahab Riaz, Hasan Ali, Mohammad Irfan, Rumman Raees, and Hassan Khan. The average age of the PCB's ICC T20 World Cup squad is 26 (as of December 2022, from ESPN Cricinfo). Considering this, nine of these ten players are above the age of 26, with six of them surpassing 30 years of age. However, it is worth noting that four players in the official squad are above 30, including Shan Masood (33), Iftikhar Ahmed (32), Fakhar Zaman (32), and Asif Ali (31). Fakhar Zaman was benched early in the tournament due to an injury and was subsequently replaced, while the remaining players did not deliver notable performances as batters. Furthermore, some players such as Mohammad Irfan and Rumman Raees have not played T20 international cricket since 2019 and 2018, respectively, which may have been a factor in their exclusion from the official squad.

Among the players in the official Pakistan squad, there were 7 players who did not appear on any of our teams. These players include batters Haider Ali, Shan Masood, and Mohammad Haris, as well as bowlers Naseem Shah, Mohammad Wasim, Shahnawaz Dahani, and Usman Qadir. However, when we exclude the first top 10 players who appear in the rankings, many of the immediate replacements align with the players selected for the tournament. For instance, all-rounders Shadab Khan ranked fourth in betweenness centrality, Mohammad Nawaz ranked fourth in both closeness and degree centrality, and Mohammad Wasim featured in the top ten in all three categories. Among the batters, Shan Masood and Asif Ali respectively ranked seventh and eighth in both closeness and degree centrality. In terms of the bowlers, Mohammad Hasnain ranked eighth in both closeness and degree centrality, while Naseem Shah secured a position in the top ten for betweenness and closeness centrality, as well as ninth in degree centrality. 

As a newcomer, Mohammad Haris emerged as a standout performer during the tournament. Initially a reserve player, he replaced Fakhar Zaman due to an injury and scored much-needed runs for his team, surpassing the achievements of many senior team members. On the other hand, the remaining players—Haider Ali, Usman Qadir, and Shahnawaz Dahani—ranked lower in our analyses and were not featured prominently in the starting 11 of many matches during the tournament. Thus, the reasons for their selection remain unclear. Interestingly, despite the presence of most of the batters in our rankings, the overall batting performance in the tournament was lackluster. The batters struggled to establish and chase target runs in numerous matches. The openers, Mohammad Rizwan and captain Babar Azam, along with the middle-order batters Khushdil Shah, Shan Masood, and Iftikhar Ahmed, ranked high in our rankings but underperformed consistently, losing their wickets early, and struggling to make significant contributions. However, there were two exceptions who stood out and ranked well in our ranking; Mohammad Haris, and all-rounder Shadab Khan. On the other hand, the bowlers delivered exceptional performances. Shaheen Afridi and Haris Rauf, who ranked high in our rankings, took numerous crucial wickets in the tournament. Additionally, Mohammad Wasim, Naseem Shah, and Shadab Khan also made significant contributions. Although Mohammad Hasnain placed lower in our rankings and did not feature in the starting 11 in several matches, his limited appearances demonstrated incredible talent. Newcomer Naseem Shah, a standout performer in both the World Cup and his PSL career, played an instrumental role by taking key wickets and frequently delivering dot balls (no runs), thereby showcasing his immense value to the team. 

\begin{table*}[ht]
\caption{Official ICC T-20 Men's World Cup Pakistan Squad}
\label{tab2}
\resizebox{\textwidth}{!}{ 
\begin{tabular}{|c|c|c|c|c|c|c|c|}
\hline
Player Name & Matches & Runs & Wickets & Degree Distribution & Betweenness Centrality & Closeness Centrality & Clustering Coefficient \\
\hline
\rule[-1ex]{0pt}{3.2ex}Fakhar Zaman (BAT)         & 63      & 1939 & 2       & 152                                     & 1952.2596                               & 0.0020                              & 0.1643                                 \\
Babar Azam (BAT)           & 68      & 2413 & 0       & 148                                     & 1385.0951                               & 0.0020                                    & 0.1958                                 \\
Iftikhar Ahmed (BAT)       & 48      & 584  & 9       & 132                                     & 1367.2141                               & 0.0019                              & 0.2074                                 \\
Khushdil Shah (BAT)        & 38      & 554  & 18      & 114                                     & 672.3614                                & 0.0019                              & 0.1988                                  \\
Shan Masood (BAT)          & 34      & 1082 & 0       & 104                                     & 579.8780                                & 0.0019                              & 0.2494                                 \\
Asif Ali (BAT)             & 67      & 1016 & 2       & 100                                     & 1042.2667                               & 0.0019                              & 0.1302                                 \\
Haider Ali (BAT)           & 28      & 557  & 0       & 76                                      & 195.7738                               & 0.0018                              & 0.2488                                 \\
Mohammad Haris (BAT)       & 5       & 166  & 0       & 38                                      & 91.2508                                 & 0.0017                              & 0.2432                                 \\
Mohammad Rizwan (WK)       & 59      & 1446 & 0       & 130                                     & 960.8386                               & 0.0019                              & 0.2053                                  \\
Shadab Khan (ALL)          & 61      & 800  & 65      & 204                                     & 2087.3407                               & 0.0021                              & 0.2263                                 \\
Mohammad Nawaz (ALL)       & 66      & 619  & 61      & 184                                     & 2105.1272                               & 0.0020                               & 0.1312                                 \\
Mohammad Wasim (ALL)       & 19      & 77   & 20      & 100                                     & 539.7155                               & 0.0019                              & 0.2334                                 \\
Shaheen Shah Afridi (BALL) & 50      & 88   & 70      & 190                                     & 2134.6579                               & 0.0021                               & 0.1917                                 \\
Haris Rauf (BALL)          & 40      & 58   & 47      & 134                                     & 1151.6622                               & 0.0019                              & 0.1968                               \\
Mohammad Hasnain (BALL)    & 28      & 8    & 39      & 106                                     & 999.5049                               & 0.0019                              & 0.2342                                 \\
Naseem Shah (BALL)         & 19      & 16   & 19      & 102                                     & 597.4772                               & 0.0018                              & 0.2438                                 \\
Shahnawaz Dhani (BALL)     & 22      & 6    & 37      & 76                                      & 462.7219                                & 0.0017                               & 0.1131                                 \\
\rule[-1ex]{0pt}{1.5ex} Usman Qadir (BALL)         & 12      & 29   & 13      & 32                                      & 52.5758                                & 0.0015                              & 0.2030  \\
\hline
\multicolumn{4}{l}{\rule[-1ex]{0pt}{3.5ex}BAT: Batter; BALL: Bowler; WK: Wicket-keeper; All: All-rounder.} \\
\end{tabular}
}
\end{table*}

\section{Conclusion \& Future Improvements}
\label{sec6}
In this paper, we utilized the approach proposed by Dey et al. \cite{b1} to assess the significance of network centrality and clustering coefficient measures as a key factor in the formation of optimal Pakistan cricket teams for international championships. Out of the 18 players in the official Pakistan squad, 11 were included in the teams selected using this approach. Although most of the remaining 7 players that were not part of our teams were still selected for the tournament, they appeared highly in our rankings. While there were some inconsistencies between player performances and selection, the majority of our top-ranked players have been observed to consistently deliver excellent performances throughout their PSL careers, according to Table \ref{tab2}, which indicates their indispensable role within the network due to their exceptional skills. Therefore, it suggests that the Pakistan Cricket Board (PCB) highly values hard work and player performance when making selections. While there is a possibility that certain players received precedence due to personal connections, these findings underscore the importance of good performance in the selection process. Hence, the obtained outcomes suggest that network centrality can serve as a feasible approach for forming optimal teams, not limited to the Pakistani context. 

Although our study did not reproduce the results reported by Dey et al. \cite{b1} concerning the local clustering coefficient measure, investigating the underlying reasons for this disparity requires more advanced network analysis measures and falls beyond the scope of this research, and can be explored for future work. Moreover, during our study, we encountered limitations, described as follows:
\begin{itemize}
    \item High centrality metrics do not provide a complete reflection of a player's ability, as certain players may have been benched due to injuries, or primarily served as reserved players, or switched teams during their PSL career. Moreover, these figures do not account the benefits associated with different roles within the game. Batters, for instance, can only perform during the batting innings if their preceding peers lose their wicket and their turn comes up. Meanwhile, all bowlers in the starting 11 are guaranteed their 5 overs to showcase their skills in the innings. Wicket-keepers and all-rounders, on the other hand, have the opportunity to contribute in both innings. These nuances are not captured in the data and should be taken into account when interpreting centrality measures as they could potentially account for certain discrepancies observed among the formed teams.
    
    \item Due to the limited availability of literature regarding players' performances, much of the analysis conducted on their form and performance over the years relied heavily on our personal observations from watching the PSL and ICC Men's T20 World Cup tournaments.
\end{itemize}
Thus, for future research, it is worth exploring temporal analysis to gain insights into the factors contributing to a cricket team's success. Analyzing changes in team squads over time can potentially reveal which players play pivotal roles, as their inclusion may have a positive impact on the team's performance. Moreover, given that the PSL is an ongoing annual tournament, incorporating more data on players' performances in upcoming matches will further enhance the rankings of new players, increasing their likelihood of being selected for our optimal teams. This will result in improved accuracy in terms of the team's resemblance to the official Pakistan squad, as it was observed to be comprised of predominantly new and young performers.

\section*{Acknowledgment}
This work was spun off from the project work in the 'Social Network Analysis' course that Abeer Khan, Maria Hunaid Samiwala, and Abeeha Zawar attended at Habib University, co-taught by Dr. Shah Jamal Alam and Dr. Qasim Pasta. 

\bibliography{references}{}
\bibliographystyle{IEEEtran}

\end{document}